\title{EnergyEfficiency_double}
\documentclass[journal]{IEEEtran}
\ifCLASSINFOpdf
\else
\fi

\usepackage{cite,times,amsmath,epsfig,algorithmic,array}
\usepackage{mdwmath}
\usepackage{mdwtab}
\usepackage{eqparbox}
\usepackage{cite}
\usepackage{graphicx}
\usepackage{amsfonts}
\usepackage{epstopdf}
\usepackage{amsfonts}
\usepackage{amsmath}
\usepackage{amssymb}
\usepackage{amsthm}
\usepackage{xcolor}
\definecolor{red}{rgb}{1.0, 0.0, 0.0}
\definecolor{blue}{rgb}{0.3,0.3,0.9}
\definecolor{darkgreen}{rgb}{0.0, 0.2, 0.13}
\definecolor{americanrose}{rgb}{1.0, 0.01, 0.24}
\definecolor{britishracinggreen}{rgb}{0.0, 0.26, 0.15}
\usepackage{algorithm}
\usepackage{algorithmic}
\RequirePackage[colorlinks,citecolor=blue,urlcolor=blue]{hyperref}

\begin{document}
%
\title{Energy Efficiency in Multiuser Transmission Over Parallel Frequency Channels}
%
%
%

\author{Liang~Dong and Xing~Meng 
\thanks{L.~Dong is with the Department
of Electrical and Computer Engineering, Baylor University, Waco, TX 76798 USA (e-mail: liang\_dong@baylor.edu).}
\thanks{X.~Meng is with the Department of Statistical Science, Baylor University, Waco, TX 76798 USA (e-mail: xing\_meng@baylor.edu).}
}

\maketitle

\begin{abstract}
Energy efficiency is an important design criterion for wireless communications.   When parallel frequency channels are used for multiuser transmission, the channel bandwidths and user power are adjusted to maximize the sum information rate with the bandwidth budget, the transmit power budget, and the user-specific rate requirements.  The maximum sum rate is used in measuring the energy efficiency.    With fixed or flexible bandwidths of the frequency channels, practical methods are developed to find the total transmit power with the unique optimal resource (bandwidth and power) allocation for maximum energy efficiency.   This resource allocation ensures that, while each user's minimum rate requirement is satisfied, all the excess resource of the spectrum and transmit power is dedicated to the one user with the best channel quality.
Simulation results validate the optimal solutions of total transmit power and resource allocation that support the energy-efficient multiuser transmission.  
\end{abstract}


%
\IEEEpeerreviewmaketitle

\section{Introduction}

Energy efficiency is an important metric of wireless communications. It is measured as the communication data rate per unit of power used. In wireless communications, a large portion of the energy is consumed for information transmission. Maximizing energy efficiency is to minimize energy consumption for transferring a certain amount of information. Energy efficiency is critical to limit excessive power usage in the communication network and to reduce the operational expenditure~\cite{6957151,7029696,7244345,7542538}.


For multiuser transmission, parallel frequency channels can be used to avoid multiuser interference.  The frequency channels with various bandwidths are assigned to and the total transmit power is split among the multiple users.  The bandwidth and power allocations are designed to maximize the sum information rate and hence improving the energy efficiency.  If the transmitter uses one of the parallel channels, it only transmits the information to one receiver with the best reception.  The optimal power allocation across the channels is a water-filling solution~\cite{612942}.   When the frequency spectrum is divided into discrete frequency bins to be assigned to different users with dedicated power spectral density, a convex programming relaxation can be exploited to find the optimal bin assignment and its energy distribution~\cite{975766,1306617}.  For practical applications in~\cite{1306617}, computational complexity is reduced by restricting the energy distribution to be constant across the used channels.  The dynamic bandwidth assignment and adaptive power allocation are referred to as resource allocation.  Over parallel Gaussian broadcast channels, the optimal resource allocation is studied to achieve any point on the boundary of the capacity region~\cite{1650370} and with a sum power constraint and the receiver-specific rate constraints~\cite{4385790}.

The resource allocation algorithms for multiuser transmission over parallel frequency channels can be applied to the orthogonal frequency division multiple access (OFDMA) systems~\cite{5208735,4595667,5164968,5272480,6762926,6824265,7036057}.  Because the energy efficiency with optimal resource allocation is strictly quasiconcave in the total transmit power, there exists a unique global maximum~\cite{6294413,6365872,6787103,7805400}.  In~\cite{6294413}, the energy efficiency is maximized under certain user quality-of-service requirements.  The optimal energy-efficient resource allocation is derived with a given subcarrier assignment, and a suboptimal algorithm is developed with a flexible subcarrier assignment.

In this paper, we address the energy-efficient transmission over parallel frequency channels.  The transmission to the multiple receivers exploits the multiuser diversity, as a different frequency is chosen for each user around which the channel response is relatively good.  This can be, for example, the downlink transmission of the contiguous OFDMA system.  Adjacent frequency bins are used for transmission to one user, and the channel responses across these bins are relatively flat and can be estimated and fed back to the transmitter.

Each user has an individual requirement of minimum date rate.    With optimal resource allocation, the sum of information rates is used in measuring the energy efficiency.  When the bandwidth of each parallel frequency channels is given, we develop a practical method to find the total transmit power and its user allocation for maximum energy efficiency.  When the bandwidths of parallel frequency channels are undetermined, we develop a practical method for energy-efficient transmission with joint optimal bandwidth assignment and power allocation.  The maximum energy efficiency is achieved, which corresponds to a total transmit power with the unique optimal resource allocation.  This resource allocation ensures that, while each user's minimum rate requirement is satisfied, all the excess resource of the spectrum and transmit power is dedicated to the one user with the best channel quality.

The remainder of the paper is organized as follows.  In Section~\ref{sec:model}, the system model is provided for multiuser transmission over parallel frequency channels.  An optimization problem is formulated to maximize the energy efficiency while satisfying the minimum rate requirements.  With fixed bandwidth assignment of the channels, the maximization of energy efficiency is discussed in Section~\ref{sec:EE} with the optimal total transmit power and power allocation.  A practical method is developed to implement the energy-efficient multiuser transmission.  With adjustable bandwidths, Section~\ref{sec:joint_optimization} gives the solution of joint bandwidth assignment and power allocation.  Furthermore, in Section~\ref{sec:combined_efficiency}, a practical method is developed to find the optimal total transmit power and resource allocation for energy-efficient multiuser transmission.  Numerical results validate the optimal solutions for energy efficiency in Section~\ref{sec:simulation} .  Finally, conclusions are drawn in Section~\ref{sec:conclusion}.


\section{System Model and Problem Formulation}
\label{sec:model}

\begin{figure}[t!]
  \centering
    \includegraphics[width=0.45\textwidth]{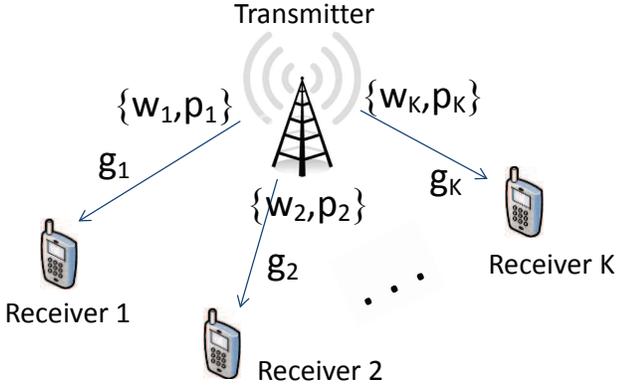}
  \caption{Transmission to multiple users over parallel frequency channels.}
    \label{fig:MAC}
\end{figure}

Consider a wireless broadcast scenario, in which a transmitter transmits to $K$ receivers in parallel frequency bands and thus there is no interference at the receivers.  Let $\mathcal{K} = \{1,2,\ldots, K\}$ be the set of the receivers.  The transmitter assigns a frequency channel of bandwidth $w_{k}$ and uses transmit power $p_{k}$ to transmit to the $k$th receiver ($ \forall k \in \mathcal{K}$).  Suppose that, once the nominal carrier frequency of the channel is determined, the bandwidth $w_{k}$ is adjustable as long as $W = \sum_{k \in \mathcal{K}} w_{k} \leq W_{M}$. $W$ is the total assigned bandwidth, and $W_{M}$ is the bandwidth budget.  An example is the contiguous subcarrier assignment of the OFDMA in IEEE 802.16 (WiMAX), IEEE 802.20 (MBWA), IEEE 802.22 (WRAN), and the downlink of the 3GPP LTE mobile broadband standard.  A group of adjacent subcarriers are assigned to a single user.  The contiguous subcarriers for one user are relatively flat over frequency, and its average channel frequency response $h_{k}$ can be measured by using pilot subcarriers.   Taking advantage of the multiuser diversity, the transmitter uses a parallel frequency channel to transmit to each receiver which avoids deep fades.  In Section~\ref{sec:joint_optimization}, we assume that $\{w_{k}\}$ are continuously adjustable.


With Gaussian channels, the achievable information rate of the $k$th user is given by
\begin{equation}
r_{k} = w_{k} \log_{2}\left( 1+ \frac{p_{k}|h_{k}|^{2}}{w_{k}N_{0}}\right) = w_{k} \log_{2}\left( 1+ \frac{p_{k} g_{k}}{w_{k}}\right)
\end{equation}
where $N_{0}$ is the noise power spectral density.   The channel gain is denoted as $g_{k} = |h_{k}|^{2}/N_{0}$ and is known to the transmitter.  It is assumed that $\{g_{k}\}$ are constant during the transmission time in consideration. Fig.~\ref{fig:MAC} illustrates the multiuser transmission.

The energy efficiency $\Gamma_{\mathrm{EE}}$ is defined as the sum achievable rate per unit of power in bps/W or bits/Joule. That is
\begin{equation}
\Gamma_{\mathrm{EE}} \triangleq \frac{R}{P_{total}} = \frac{R}{P/\zeta + P_{C}}.
\end{equation}
where $R = \sum_{k \in \mathcal{K}} r_{k}$, $P$ is the transmit power, $\zeta$ is the power amplifier efficiency to generate transmit power $P$, and $P_{C}$ is the circuit power.  Here,
$P = \sum_{k \in \mathcal{K}} p_{k} \leq P_{M}$, and
$P_{M}$ is the transmit power budget.  


The optimization of energy-efficient transmission is to maximize $\Gamma_{\mathrm{EE}}$ over $W$, $P$, and appropriately assigned $\{w_{k}\}_{k \in \mathcal{K}}$ and allocated $\{p_{k}\}_{k \in \mathcal{K}}$.  It can be formulated as
\begin{eqnarray*}
\mathcal{P}_{1}:
\begin{array}{ll}
\begin{split}
\mathop{\mathrm{maximize}}_{W, P, \{w_{k}\}, \{p_{k}\} }
\end{split}  
& \Gamma_{\mathrm{EE}} \\
\mathrm{subject~to} & r_{k} = w_{k}\log_{2}\left(1 + \frac{p_{k}g_{k}}{w_{k}} \right) \geq \check{r}_{k},  \forall k \in \mathcal{K} \\
& \sum_{k \in \mathcal{K}} w_{k} = W \leq W_{M} \\
& \sum_{k \in \mathcal{K}} p_{k} = P \leq P_{M}.
\end{array}
\end{eqnarray*}
where $\check{r}_{k}$ is the minimum achievable information rate.  It is required that $r_{k} \geq \check{r}_{k}$ $( \forall k \in \mathcal{K})$ to guarantee the quality of service.  Let $R_{0} = \sum_{k \in \mathcal{K}} \check{r}_{k}$.

We approach Problem $\mathcal{P}_1$ with a two-step progression.   First, suppose that $\{w_{k}\}$ are fixed.  For any arbitrary $P$, we derive the optimal power allocation $\{p_{k}\}$ that maximizes $R$. Further, the optimal $P$ is found with a practical method that maximizes $\Gamma_{\mathrm{EE}}$.  Second, we address the joint bandwidth assignment and transmit power allocation that maximizes $R$ and further find the optimal $P$ that maximizes $\Gamma_{\mathrm{EE}}$.

\section{Transmit Power Allocation and Energy Efficiency}
\label{sec:EE}

\subsection{Optimal Transmit Power Allocation}
\label{sec:power_allocation}

Suppose that the total bandwidth $W$ and its assignment $\{w_{k}\}_{k \in \mathcal{K}}$ for transmission to the $K$ receivers are fixed.  Given a total transmit power $P \leq P_{M}$ that is used by the transmitter, the optimization problem of transmit power allocation can be formulated as
\begin{eqnarray*}
\mathcal{P}_{2}:
\begin{array}{ll}
\begin{split}
\mathop{\mathrm{maximize}}_{\{p_{k}\}}
\end{split}  
& R =  \sum_{k \in \mathcal{K}} w_{k} \log_{2} \left(1 + \frac{p_{k}g_{k}}{w_{k}} \right) \\
\mathrm{subject~to} & p_{k} \geq \frac{w_{k}}{g_{k}} \left( 2^{\frac{\check{r}_{k}}{w_{k}}} -1 \right), \,\,\, \forall k \in \mathcal{K}\\
& \sum_{k \in \mathcal{K}} p_{k} = P
\end{array}
\end{eqnarray*}
where the constraints are to satisfy the minimum rate requirements and the total power limit.


The achievable information rate $r_{k} = w_{k} \log_{2}(1 + p_{k}g_{k}/w_{k})$ is strictly increasing and concave in $p_{k}$, for any given positive $g_{k}$ and $w_{k}$.  Problem $\mathcal{P}_{2}$ is a convex optimization problem.  The solution of $\mathcal{P}_{2}$ and its property are summarized in Proposition 1.

\textit{Proposition 1:}  
For a total transmit power $P$ that is used, the power allocation problem $\mathcal{P}_{2}$ is a convex optimization problem with its optimal solution given by
\begin{equation} \label{eq:opt_powerallocation}
\hat{p}_{k} = \check{p}_{k}+ w_{k}\left( \frac{1}{\mu} - \alpha_{k} \right)^{+}, \,\,\, \forall k \in \mathcal{K} 
\end{equation}
where
\begin{eqnarray*}
\check{p}_{k} &=& \frac{w_{k}}{g_{k}} \left( 2^{\frac{\check{r}_{k}}{w_{k}}} -1 \right) \\
\alpha_{k} &=& \frac{1}{g_{k}} + \frac{\check{p}_{k}}{w_{k}}
\end{eqnarray*}
and $(x)^{+}$ represents $\max(x,0)$ and $\mu$ is an intermediate variable that makes $\hat{p}_{k}$'s satisfy the total transmit power constraint.  

Define Set $\mathcal{I}$ as the set that contains the users transmitting with their minimum required power, i.e., $\hat{p}_{i} = \check{p}_{i}, \forall i \in \mathcal{I} \subseteq \mathcal{K}$.  The maximum sum information rate is given by
\begin{equation} \label{eq:max_EE}
\hat{R}  =   \sum_{i \in \mathcal{I} }\check{r}_{i} + \sum_{j \in \mathcal{K} \setminus \mathcal{I}} w_{j} \log_{2} \left(\frac{g_{j}}{\mu}\right) 
\end{equation}
where
\begin{eqnarray*}
\frac{1}{\mu} &=& \frac{P - P_{0} + \sum_{j \in \mathcal{K} \setminus \mathcal{I}}  w_{j} \alpha_{j}}{\sum_{j \in \mathcal{K} \setminus \mathcal{I}}w_{j}} \\
P_{0} &=& \sum_{k \in \mathcal{K}} \check{p}_{k}.
\end{eqnarray*}


~~~\textit{Proof:} See Appendix \ref{ap:proposition1}.    $\square$

\begin{figure}[t!]
  \centering
    \includegraphics[width=0.48\textwidth]{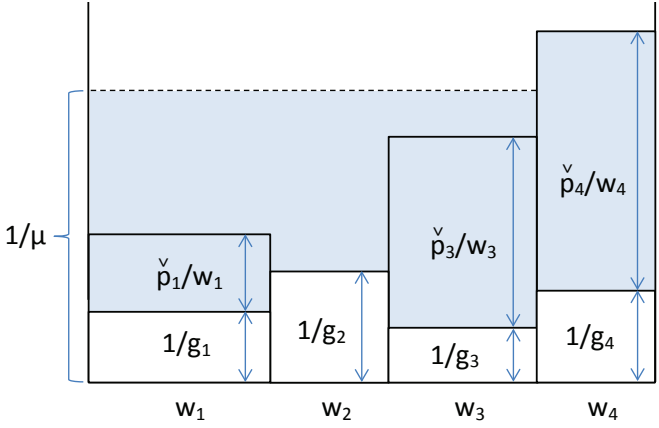}
  \caption{Water-filling illustration of optimal transmit power allocation.}
    \label{fig:waterfilling2}
\end{figure}

Given $P$, the optimal power allocation $\{\hat{p}_{k}\}$ is the water-filling solution.  This can be visualized in Fig.~\ref{fig:waterfilling2}.  The $k$th user occupies a power area with width $w_{k}$.  The initial water level is $\alpha_{k} = 1/g_{k} + \check{p}_{k}/w_{k}$.  The total transmit power $P$ is the shaded area in the figure.  When the total power $P > P_{0}$, the power water level raises to $1/\mu$. 
In this example in Fig.~\ref{fig:waterfilling2}, $\mathcal{I} = \{4\}$, $1/\mu < \alpha_{i}, \forall i \in \mathcal{I}$ and $1/\mu \geq \alpha_{j}, \forall j \in \mathcal{K} \setminus \mathcal{I}$.

\subsection{Maximization of Energy Efficiency}

With total used power $P \leq P_{M}$ and the optimal power allocation, the maximum sum information rate and the energy efficiency are $\hat{R}(P)$ and $\hat{\Gamma}_{\mathrm{EE}}(P) = \hat{R}/(P/\zeta + P_{C})$, respectively.  The goal is to find $P^{opt} \in [P_{0}, P_{M}]$ such that $\hat{\Gamma}_{\mathrm{EE}}(P^{opt})$ is maximized while applying the optimal power allocation.

\textit{Proposition 2:} The maximum sum information rate $\hat{R}(P)$ is continuously differentiable, strictly increasing, and concave in $P \geq P_{0}$.  The energy efficiency $\hat{\Gamma}_{\mathrm{EE}}(P)$ is continuously differentiable, and it is either strictly decreasing or strictly quasiconcave in $P \geq P_{0}$.  

~~~\textit{Proof:} See Appendix \ref{ap:proposition2}.    $\square$

If $\hat{\Gamma}_{\mathrm{EE}}(P)$ is strictly decreasing, the maximum energy efficiency is achieved while transmitting with minimum power for every user that meets the rate requirement, i.e., $P^{opt} = P_{0}$ and $\hat{p}_{k} = \check{p}_{k}, \forall k \in \mathcal{K}$.
If $\hat{\Gamma}_{\mathrm{EE}}(P)$ is strictly quasiconcave, since $\lim_{P \rightarrow \infty} \hat{\Gamma}_{\mathrm{EE}}(P) = 0$, there is a unique global maximum at some finite transmit power $P^{opt}$.  As $\hat{\Gamma}_{\mathrm{EE}}(P)$ is continuously differentiable, its first derivative can be calculated and used to find $P^{opt}$ that maximize the energy efficiency.  The derivative of $\hat{\Gamma}_{\mathrm{EE}}(P)$ is given by 
\begin{eqnarray}
\frac{d \hat{\Gamma}_{\mathrm{EE}}(P)}{dP} &=& \frac{d}{dP}\left( \frac{\hat{R}(P)}{P/\zeta + P_{C}}\right) \nonumber \\
&=& \frac{d \hat{R}(P)}{dP} \frac{1}{P/\zeta + P_{C}} - \frac{\hat{R}(P)}{\zeta (P/\zeta+P_{C})^{2}}.
\end{eqnarray}
As $P/\zeta + P_{C} > 0$, the sign of $d \hat{\Gamma}_{\mathrm{EE}}(P) / dP$ is determined by the sign of $\Lambda(P) = (P/\zeta +P_{C}) \cdot d \hat{\Gamma}_{\mathrm{EE}}(P) / dP$ as
\begin{equation} \label{eq:LambdaP}
\Lambda(P) = \frac{d \hat{R}(P)}{dP} - \frac{1}{\zeta}\hat{\Gamma}_{\mathrm{EE}}(P).
\end{equation}

\subsection{Practical Method}
\label{sec:EE_practical}

To implement the transmission with maximum energy efficiency, we develop a method that
can be applied to practical systems.  As total transmit power $P$ increases from $P_{0}$, the critical levels of $P$ at which Set $\mathcal{I}$ changes are found as follows.  Let $\{\alpha_{k}\}_{k = 1}^{K}$ be sorted in the ascending order and denoted as $\alpha_{1}' \leq \alpha_{2}' \leq \cdots \leq \alpha_{K}'$, where $\alpha_{k}'$ corresponds to $g_{k}'$, $\check{p}_{k}'$ and $w_{k}'$. 
According to the power allocation water-filling solution, as $P$ increases from $P_{0}$, the critical levels of $P$ are the ones at which the water level $1/\mu$ reaches $\alpha_{1}', \alpha_{2}', \ldots, \alpha_{K}'$ in this order.  In an ascending order, the critical levels are given by
\begin{equation} \label{eq:critical_power}
P_{J-1} = P_{0}+ \sum_{i=1}^{J-1}(\alpha_{J}'-\alpha_{i}')w_{i}', \,\,\, J = 1, 2, \ldots, K.
\end{equation}

According to \eqref{eq:max_EE}, at the critical level $P_{J-1}$, the energy efficiency is given by
\begin{equation} \label{eq:EE_criticallevel}
\hat{\Gamma}_{\mathrm{EE}}(P_{J-1}) = \frac{R_{0} + \sum_{i=1}^{J-1}w_{i}' \log_{2} (\alpha_{J}'/{\alpha_{i}'})}{\left(P_{0}+ \sum_{i=1}^{J-1}(\alpha_{J}'-\alpha_{i}')w_{i}'\right)/\zeta + P_{C}}.
\end{equation}
From \eqref{eq:dR_dP}, we have $d\hat{R}(P)/dP |_{P = P_{J-1}} = (1/\ln 2) (1/\alpha_{J}')$.
Let $\Lambda_{J-1}$ denote  $\Lambda(P_{J-1})$, and it follows from \eqref{eq:LambdaP} that
\begin{equation} \label{eq:Lambda_criticallevel}
\Lambda_{J-1} = \frac{1}{\ln 2} \frac{1}{\alpha_{J}'}  - \frac{1}{\zeta} \hat{\Gamma}_{\mathrm{EE}}(P_{J-1}).
\end{equation}
If $\Lambda_{0} \leq 0$, the optimal total transmit power to maximize energy efficiency is $P^{opt} = P_{0}$ with power allocation $\hat{p}_{k} = \check{p}_{k}, \forall k \in \mathcal{K}$.  In the situation when $\Lambda_{J-1} > 0$ and $\Lambda_{J} \leq 0$, $J$ users with the $J$ smallest $\alpha$'s use excess transmit power beyond their minimum required power to achieve the maximum energy efficiency, and $P^{opt} \in (P_{J-1}, P_{J}]$.  Within this power interval,  the energy efficiency with optimal power allocation can be calculated from \eqref{eq:max_EE} as
\begin{eqnarray} \label{eq:max_EE_oneinterval}
\hat{\Gamma}_{\mathrm{EE}}(P) &=& \frac{1}{P/\zeta + P_{C}}\left( A + \tilde{W}_{J} \log_{2}(P-B) \right), \nonumber \\  && \,\,\, P \in (P_{J-1}, P_{J}]
\end{eqnarray}
where
\begin{eqnarray*}
\tilde{W}_{J} &=& \sum_{i = 1}^{J} w_{i}' \\
A &=& R_{0} - \sum_{i=1}^{J} w_{i}' \log_{2}\alpha_{i}' - \tilde{W}_{J} \log_{2} \tilde{W}_{J} \\
B &=& P_{0} - \sum_{i=1}^{J} w_{i}' \alpha_{i}'. 
\end{eqnarray*}

The derivative of $\hat{\Gamma}_{\mathrm{EE}}(P)$ in this interval is given by
\begin{eqnarray}
\frac{d \hat{\Gamma}_{\mathrm{EE}}(P)}{dP} &=& \frac{\tilde{W}_{J}}{(P/\zeta + P_{C})(P-B) \ln 2} \nonumber \\
&& - \frac{A + \tilde{W}_{J}\log_{2}(P-B)}{ \zeta (P/\zeta + P_{C})^{2}}.
\end{eqnarray}
The sign of the derivative is determined by the sign of $\Theta(P) =  \zeta (P/\zeta + P_{C})^{2} \ln 2 \cdot (d \hat{\Gamma}_{\mathrm{EE}}(P)/dP) $, which is given by
\begin{equation} \label{eq:Popt_oneinterval}
\Theta(P) = \frac{(P + \zeta P_{C}) \tilde{W}_{J}}{P-B} - A \ln 2 - \tilde{W}_{J} \ln(P-B).
\end{equation}
As $\Theta(P_{J-1}) > 0$ and $\Theta(P_{J}) \leq 0$, the root of $\Theta(P), P \in (P_{J-1}, P_{J}]$ can be found with the bisection method.


The procedure of finding the optimal total transmit power $P^{opt}$ along with its user allocation $\{\hat{p}_{k}\}$ that maximizes $\Gamma_{\mathrm{EE}}$  is sketched in Algorithm~\ref{alg:power}.

\begin{algorithm}[h]
\caption{Find optimal total transmit power and its user allocation that maximize $\Gamma_{\mathrm{EE}}$}\label{alg:power}
\begin{algorithmic}
\item[0.] Calculate minimum transmit power $\check{p}_{k}$ from minimum rate requirement $\check{r}_{k}, \forall k \in \mathcal{K}$;
\item[1.] Calculate $\alpha_{k} = 1/g_{k} + \check{p}_{k}/w_{k}$, where $g_{k}, w_{k}, \forall k \in \mathcal{K}$ are given;
\item[2.] Sort $\{\alpha_{k}\}$ in ascending order as $\{\alpha_{k}'\}_{k=1}^{K}$;
\item[3.] Calculate the critical levels of total transmit power $P_{J-1}, J = 1, 2, \ldots, K$ using \eqref{eq:critical_power}.  The power levels can be neglected if they are beyond $P_{M}$;
\item[4.] Calculate $\hat{\Gamma}_{\mathrm{EE}}(P_{J-1})$ using \eqref{eq:EE_criticallevel} and $\Lambda_{J-1}$ using \eqref{eq:Lambda_criticallevel}, starting with $J=1$ and stopping when $\Lambda$ is negative;
\item[5.] If $\Lambda_{0} \leq 0$, $P^{opt} = P_{0}$, go to Step 7;
\item[6.] When $\Lambda_{J-1} > 0$  and $\Lambda_{J} \leq 0$, the optimal total transmit power is in interval $(P_{J-1} , P_{J}]$. Establish Set $\mathcal{I}$.  Find $P^{opt}$ as the root of $\Theta(P)$ in \eqref{eq:Popt_oneinterval} using the bisection method;
\item[7.] With $P^{opt}$, calculate the optimal transmit power allocation using \eqref{eq:opt_powerallocation} and the maximum energy efficiency using \eqref{eq:max_EE_oneinterval}.
\end{algorithmic}
\end{algorithm}

\section{Joint Bandwidth Assignment and Transmit Power Allocation}
\label{sec:joint_optimization}

With a total bandwidth $W$ and total used transmit power $P$, what is the best way to simultaneously assign frequency channel bandwidths and allocate power for transmission to the $K$ users?  Suppose that the transmitter can exploit the multiuser diversity and determine different nominal carrier frequencies for transmission to the appropriate receivers.   The channel bandwidths $\{w_{k}\}$ of the parallel frequency bands are continuously adjustable~\cite{1427691}.   Within the $k$th frequency band, suppose that the channel frequency response is flat and the channel gain $g_{k}$ can be measured.

As the criterion is to maximize the sum information rate $R$, the optimization problem of joint bandwidth assignment and transmit power allocation can be written as
\begin{eqnarray*}
\mathcal{P}_{3}:
\begin{array}{ll}
\begin{split}
\mathop{\mathrm{maximize}}_{\{w_{k}\}, \{p_{k}\} }
\end{split}  
& R = \sum_{k \in \mathcal{K}} w_{k} \log_{2} \left(1 + \frac{p_{k}g_{k}}{w_{k}} \right) \\
\mathrm{subject~to} & w_{k}\log_{2}\left(1 + \frac{p_{k}g_{k}}{w_{k}} \right) \geq \check{r}_{k}, \,\,\, \forall k \in \mathcal{K} \\
& \sum_{k \in \mathcal{K} } w_{k} = W \\
& \sum_{k \in \mathcal{K} } p_{k} = P .
\end{array}
\end{eqnarray*}
Suppose that $\{g_{k}\}$ are distinct and are sorted in the descending order $g_{1} > g_{2} > \cdots > g_{K}$.  The sum information rate $R$ is a concave function of $\{w_{k}\}$ and $\{p_{k}\}$.  Therefore, 
Problem $\mathcal{P}_{3}$ is a convex optimization problem.   

\textit{Proposition 3:}  
For a total bandwidth $W$ and total used transmit power $P$, problem $\mathcal{P}_{3}$ of joint bandwidth assignment and transmit power allocation is a convex optimization problem.  If this problem has a feasible solution, the optimum is achieved as follows.
\begin{enumerate}
  \item The $K-1$ users with channel gains $g_{i}, i = 2,3, \ldots, K$, are transmitted to at their minimum rate $\check{r}_{i}, i = 2, 3, \ldots, K$.  All of the remaining resources of the spectrum and the transmit power is used for transmission to the user with the best channel gain $g_{1}$.
  \item The bandwidth assigned and the transmit power allocated to the $i$th user ($i = 2, 3, \ldots, K$) are respectively
  \begin{eqnarray}
w_{i} &=& \frac{\check{r}_{i} \ln 2}{\mathcal{W}_{0}^{(i)} + 1} \label{eq:w_i}\\
p_{i} &=& \frac{\check{r}_{i} \ln 2}{g_{i}} \cdot  \frac{\psi g_{i}-1-\mathcal{W}_{0}^{(i)}}{\mathcal{W}_{0}^{(i)}(\mathcal{W}_{0}^{(i)} + 1)}  \label{eq:p_i} 
\end{eqnarray}
where $\mathcal{W}_{0}^{(i)} = \mathcal{W}_{0}(\frac{\psi g_{i}-1}{e})$, and $\mathcal{W}_{0}(\cdot)$ is the principal branch of the Lambert W function.  \\The bandwidth assigned and the transmit power allocated to the user with $g_{1}$ are respectively
\begin{eqnarray} 
w_{1} &=& W - \sum_{i=2}^{K}w_{i} \label{eq:w1_joint} \\
p_{1} &=& P - \sum_{i=2}^{K}p_{i}. \label{eq:p1_joint}
\end{eqnarray}
The intermediate coefficient $\psi$ satisfies
\begin{equation}
\psi = \left(\frac{1}{g_{1}} + \frac{p_{1}}{w_{1}} \right) \ln\left(1 + \frac{p_{1}g_{1}}{w_{1}} \right) - \frac{p_{1}}{w_{1}}.  \label{eq:phi}
\end{equation}
  \item The maximum sum information rate is
  \begin{equation}
\hat{R} = \sum_{i = 2}^{K}\check{r}_{i} + \frac{w_{1}}{\ln 2} \left( \mathcal{W}_{0}^{(1)} + 1\right)   \label{eq:max_rate_joint}
\end{equation}
where $\mathcal{W}_{0}^{(1)} = \mathcal{W}_{0}(\frac{\psi g_{1}-1}{e})$.
\end{enumerate}

~~~\textit{Proof:} See Appendix \ref{ap:proposition3}.    $\square$

For $i = 2, 3, \ldots, K$, it follows \eqref{eq:w_i} that  $w_{i}$ is strictly decreasing in $\psi > 0$.  Therefore, $p_{i}$ is strictly increasing in $\psi > 0$.  From \eqref{eq:p_i}, we have $\lim_{\psi \rightarrow 0} p_{i} = \check{r}_{i} \ln 2 / g_{i}, i = 2,3,\ldots,K$.   

For $i=1$, the maximum information rate of the user with channel gain $g_{1}$ is denoted as $\hat{r}_{1}$, i.e., $\hat{r}_{1} = w_{1}\log_{2}(1 + p_{1}g_{1}/w_{1})$, and
\eqref{eq:phi} can be rewritten as
\begin{equation}
\psi = \frac{1}{g_{1}} \left(2^{\frac{\hat{r}_{1}}{w_{1}}} \left( \frac{\hat{r}_{1}}{w_{1}} \ln 2 -1 \right) +1 \right).
\end{equation}
The bandwidth assigned to this user can be solved as
\begin{equation}
w_{1} = \frac{\hat{r}_{1} \ln 2}{\mathcal{W}_{0}^{(1)} + 1}.  \label{eq:w1}
\end{equation}
Let us denote $\check{w}_{1} = \check{r}_{1} \ln 2 / (\mathcal{W}_{0}^{(1)}+1)$ and $\check{p}_{1}$ as the bandwidth and the transmit power, respectively, that just meet the minimum rate requirement $\check{r}_{1}$ of user 1.  The formula of $\check{w}_{1}$ is analogous to \eqref{eq:w_i} which shows that $w$ is strictly decreasing in $\psi$.
Because $\hat{r}_{1} \geq \check{r}_{1}$, the feasible value of $\psi$ is in the interval $\psi \in [\psi_{\mathrm{min}}, \psi_{\mathrm{max}}]$, where  $\psi_{\mathrm{min}}$ is found when $\check{w}_{1}+\sum_{i=2}^{K}w_{i} = W$ and $\psi_{\mathrm{max}}$ is found when $\check{p}_{1} + \sum_{i=2}^{K}p_{i} = P$.

If $\psi_{\mathrm{min}} > \psi_{\mathrm{max}}$, the minimum rate requirements $\{\check{r}_{k}\}_{k \in \mathcal{K}}$ cannot be simultaneously satisfied with the total bandwidth $W$ and the total transmit power $P$.  
For $\psi_{\mathrm{min}} \leq \psi_{\mathrm{max}}$, it follows from \eqref{eq:w1} that
\begin{equation}
\mathcal{W}_{0}^{(1)} + 1 = \ln \left(1 + \frac{(P-\sum_{i=2}^{K}p_{i}) g_{1}}{W - \sum_{i=2}^{K}w_{i}} \right). \label{eq:find_psi}
\end{equation}
The left side of \eqref{eq:find_psi} is strictly increasing and the right side of \eqref{eq:find_psi} is strictly decreasing in $\psi \in [\psi_{\mathrm{min}}, \psi_{\mathrm{max}}]$.  Define $\Psi(\psi)$ as
\begin{equation} \label{eq:Psi}
\Psi(\psi) = \mathcal{W}_{0}^{(1)} + 1 - \ln \left(1 + \frac{(P-\sum_{i=2}^{K}p_{i}) g_{1}}{W - \sum_{i=2}^{K}w_{i}} \right).
\end{equation}
It has a unique zero-crossing point in $[\psi_{\mathrm{min}}, \psi_{\mathrm{max}}]$, and $\psi$ can be obtained using numerical methods.

\section{Energy Efficiency in Multiuser Transmission}
\label{sec:combined_efficiency}

The total bandwidth $W$ and the total transmit power $P$ are to be adjusted to maximize the energy efficiency $\hat{\Gamma}_{\mathrm{EE}} = \hat{R}(W,P)/(P/\zeta + P_{C})$.  Of course, with every particular $W$ and $P$, the maximum sum information rate $\hat{R}(W,P)$ is achieved with the proposed method of joint bandwidth assignment and transmit power allocation.

\textit{Proposition 4:} The energy efficiency $\hat{\Gamma}_{\mathrm{EE}} = \left(\frac{1}{P/\zeta + P_{C}} \right) \hat{R}(W,P)$  is continuously differentiable in $W$ and $P$, given that the $(W, P)$-pair can satisfy the minimum rate requirements.   The energy efficiency $\hat{\Gamma}_{\mathrm{EE}}$ is strictly increasing and concave in $W$, and $\hat{\Gamma}_{\mathrm{EE}}$ is either strictly decreasing or strictly quasiconcave in $P$. 

~~~\textit{Proof:} The proof is similar to the proof of Proposition 2, starting with that $\hat{R}(W,P)$ is strictly increasing and concave in $W$ and $P$.    $\square$


It is evident from Proposition 4 that the multiuser transmission should use total bandwidth as wide as possible, i.e., $W = W_{M}$.

If the total bandwidth $W = W_{M}$ is small, sufficient transmit power $P_{0}$ is required to meet the minimum rate requirements.  In this case, it is more likely that $\hat{\Gamma}_{\mathrm{EE}}$ is strictly decreasing in $P$.    If $W = W_{M}$ is large, there is already adequate bandwidth that can be used to meet the minimum rate requirements.  In this case, it is more likely that $\Gamma_{\mathrm{EE}}$ is quasiconcave in $P$.  The derivative-based method can be used to find the optimal total bandwidth $P^{opt}$ that maximizes $\hat{\Gamma}_{\mathrm{EE}}$.
The partial derivative of the energy efficiency over $P$ is given by
\begin{IEEEeqnarray}{lCl}
\lefteqn{\frac{\partial \hat{\Gamma}_{\mathrm{EE}}(P)}{\partial P} } \nonumber \\ 
&=& 
\frac{\partial}{\partial P} \left[ \left(\frac{1}{P/\zeta +P_{C}} \right) \hat{R}(W,P) \right] \nonumber \\
&=& \frac{1}{P/\zeta + P_{C}} \frac{\partial \hat{R}(W,P)}{\partial P} - \frac{\hat{R}(W,P)}{\zeta (P/\zeta + P_{C})^{2}} \nonumber \\
&=& \frac{1}{P/\zeta +P_{C}} \frac{\partial}{\partial P} \left[ \left(\frac{W}{\ln 2} - \sum_{i=2}^{K}\frac{\check{r}_{i}}{\mathcal{W}_{0}^{(i)} + 1} \right)  \left( \mathcal{W}_{0}^{(1)} + 1 \right) \right] \nonumber \\
&& -\frac{1}{\zeta (P/\zeta + P_{C})^{2}} \left( \sum_{i = 2}^{K} \check{r}_{i} + \frac{w_{1}}{\ln 2} \left( \mathcal{W}_{0}^{(1)} + 1\right) \right) \nonumber \\
&=& \frac{1}{P/\zeta + P_{C}}  \left[ \Omega_{1} \sum_{i=2}^{K} \check{r}_{i} \Omega_{i}^{-2} \frac{g_{i}}{\psi g_{i}-1 + \exp( \Omega_{i})}  \right. \nonumber \\
&& \left. + \left(\frac{W}{\ln 2} - \sum_{i=2}^{K} \check{r}_{i} \Omega_{i}^{-1} \right) \frac{g_{1}}{\psi g_{1} -1 + \exp(\Omega_{1})} \right]  \frac{\partial \psi}{\partial P} \nonumber \\
&&  -\frac{1}{\zeta (P/\zeta + P_{C})^{2}} \left( \sum_{i = 2}^{K} \check{r}_{i} + \left(\frac{W}{\ln 2} - \sum_{i=2}^{K} \check{r}_{i} \Omega_{i}^{-1} \right) \Omega_{1} \right) \nonumber \\
&=& \frac{1}{P/\zeta + P_{C}} F_{1}(W,P) \frac{\partial \psi}{\partial P } - \frac{1}{\zeta (P/\zeta + P_{C})^{2}} F_{2}(W,P) \nonumber \\
\end{IEEEeqnarray}
where $\psi(W,P)$ is obtained following Proposition 3.  
For particular $W$ and $P$ that can satisfy the minimum rate requirements, there is a unique $\psi$ that corresponds to $\hat{R}(W,P)$ in \eqref{eq:max_rate_joint}.  As $P$ changes, $\psi$ changes.  With $\psi(W,P)$ obtained for a series of $P$, its partial derivative $\partial \psi / \partial P$ can be calculated. And,
\begin{eqnarray*}
\Omega_{i} &=& \mathcal{W}_{0}^{(i)}  +1, \,\,\, i = 1, 2, \ldots, K \\
F_{1}(W,P) &=& \Omega_{1} \sum_{i=2}^{K} \check{r}_{i} \Omega_{i}^{-2} \frac{g_{i}}{\psi g_{i}-1 + \exp( \Omega_{i})} \nonumber \\
&& + \left(\frac{W}{\ln 2} - \sum_{i=2}^{K} \check{r}_{i} \Omega_{i}^{-1} \right) \frac{g_{1}}{\psi g_{1} -1 + \exp(\Omega_{1})} \nonumber \\
F_{2}(W,P) &=& \sum_{i = 2}^{K} \check{r}_{i} + \left(\frac{W}{\ln 2} - \sum_{i=2}^{K} \check{r}_{i} \Omega_{i}^{-1} \right) \Omega_{1}. \nonumber
\end{eqnarray*}

The sign of the partial derivative of the energy efficiency over $P$ is determined by the sign of $\Lambda_{P}(W,P)$, which is defined as
\begin{equation} \label{eq:Lambda}
\Lambda_{P}(W,P) = (P + \zeta P_{C}) F_{1}(W,P) \frac{\partial \psi}{\partial P} - F_{2}(W,P).
\end{equation}
If $\hat{\Gamma}_{\mathrm{EE}}$ is strictly decreasing in $P$ and $\Lambda_{P} <0$ ($ \forall P \leq P_{M}$), the optimal total transmit power $P^{opt}$ is the minimum required transmit power, i.e., $P^{opt} = P_{0}$.  If $\hat{\Gamma}_{\mathrm{EE}}$ is quasiconcave, $P^{opt}$ is found at the zero-crossing point $\Lambda_{P}(W,P^{opt}) = 0$.


The procedure of maximizing the energy efficiency $\hat{\Gamma}_{\mathrm{EE}}$ in multiuser transmission  is sketched in Algorithm~\ref{alg:joint}.

\begin{algorithm}[h]
\caption{Find $P^{opt}$ that maximizes energy efficiency $\hat{\Gamma}_{\mathrm{EE}}$ with joint bandwidth assignment and transmit power allocation}\label{alg:joint}
\begin{algorithmic}
\item[0.] Different nominal carrier frequencies are chosen for transmission to the $K$ receivers such that each user has a relatively good channel gain.    Sort the channel gains in a descending order $g_{1} > g_{2} > \cdots > g_{K}$;
\item[1.] Total available bandwidth $W_{M}$ is used to assign parallel frequency channels to the $K$ users. Find $\psi_{\mathrm{min}}$ by setting $\check{w}_{1}+\sum_{i=2}^{K}w_{i} = W_{M}$, where \eqref{eq:w_i} is used for $\{w_{i}\}$;
\item[2.] With $\psi = \psi_{\mathrm{min}}$, calculate $P_{0} = \check{p}_{1} + \sum_{i=2}^{K}p_{i}$, where \eqref{eq:p_i} is used for $\{p_{i}\}$.   If $P_{0} > P_{M}$, the minimum rate requirements cannot be met simultaneously, and Stop;
\item[3.] For $P \in [P_{0}, P_{M}]$, find $\psi(P)$ such that $\Psi(\psi(P)) = 0$ in \eqref{eq:Psi}.  These values can be obtained with simple numerical methods because $\Psi(\psi)$ is monotonically increasing in $\psi$. Calculate $\partial \psi(P) / \partial P$;
\item[4.] For $P \in [P_{0}, P_{M}]$, calculate $F_{1}(W_{M}, P)$ and $F_{2}(W_{M}, P)$.  Calculate $\Lambda_{P}(W_{M}, P)$ using \eqref{eq:Lambda};
\item[5.] If $\Lambda_{P}(W_{M}, P) < 0$, $\forall P \in [P_{0}, P_{M}]$, $P^{opt} = P_{0}$.  Otherwise, $P^{opt}$ is found when $\Lambda_{P}(W_{M}, P^{opt}) = 0$.  Simple numerical methods can be used because $\Lambda_{P}(W_{M}, P)$ is monotonically decreasing in $P$;
\item[6.] The maximum energy efficiency is $\hat{\Gamma}_{\mathrm{EE}} = \hat{R}(W_{M}, P^{opt})/(P^{opt}/\zeta + P_{C})$, where $\hat{R}$ is calculated using \eqref{eq:max_rate_joint}.  With $W = W_{M}$ and $P = P^{opt}$, the joint bandwidth assignment and transmit power allocation are obtained with \eqref{eq:w_i}--\eqref{eq:p1_joint}.
\end{algorithmic}
\end{algorithm}

\section{Numerical Results}
\label{sec:simulation}

We simulate a scenario in which a transmitter communicates with $K=3$ receivers over parallel frequency channels.   The channel gains $\{g_{k}\}$ are arbitrary but fixed during the transmission.  The three users have minimum rate requirements $\{\check{r}_{k}\}$.

First, suppose that the bandwidth assigned to the users $\{w_{k}\}$ are fixed.  We randomly generate $g_{k}$ from $(0, 10]$, 
randomly generate $w_{k}$ such that $W$ is in $(0, 15]$, and calculate $\{\check{p}_{k}\}$ from $\{\check{r}_{k}\}$.  The total transmit power $P$ increases starting from $P_{0}$.   Given $P$, the transmit power allocation $\{p_{k}\}$ are found according to the proposed method in Section~\ref{sec:power_allocation}.  Suppose that $\zeta = 0.8$ and $P_{C} = 10$.

\begin{figure}[t!]
  \centering
    \includegraphics[width=0.53\textwidth]{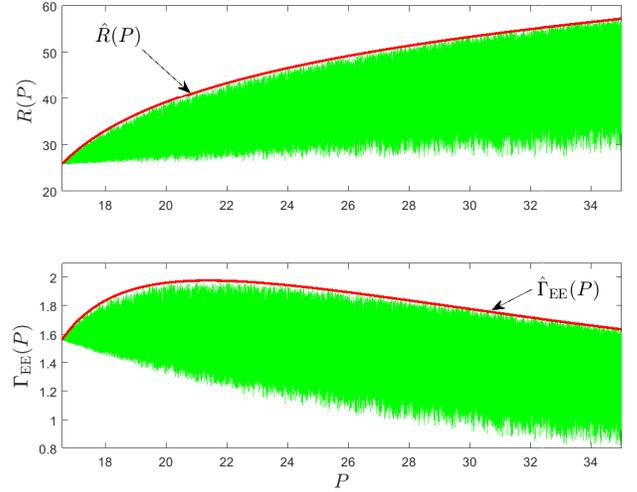}
  \caption{Sum information rate and energy efficiency with transmit power allocations.  Fixed bandwidth assignment.}
    \label{fig:EE_optsolution}
\end{figure}

The top figure of Fig.~\ref{fig:EE_optsolution} shows the sum information rate $R(P)$ versus $P$.  The green area contains
$R(P)$ with various power allocations.  The solid red curve represents the sum rate $\hat{R}(P)$ with the optimal transmit power allocation.  The bottom figure of Fig.~\ref{fig:EE_optsolution} shows the energy efficiency $\Gamma_\mathrm{EE}(P)$ versus $P$.   The green area contains $\Gamma_\mathrm{EE}(P)$ with various power allocations.  The solid red curve represents $\hat{\Gamma}_\mathrm{EE}(P)$ with the optimal transmit power allocation, which is quasiconcave.

\begin{figure}[t!]
  \centering
    \includegraphics[width=0.48\textwidth]{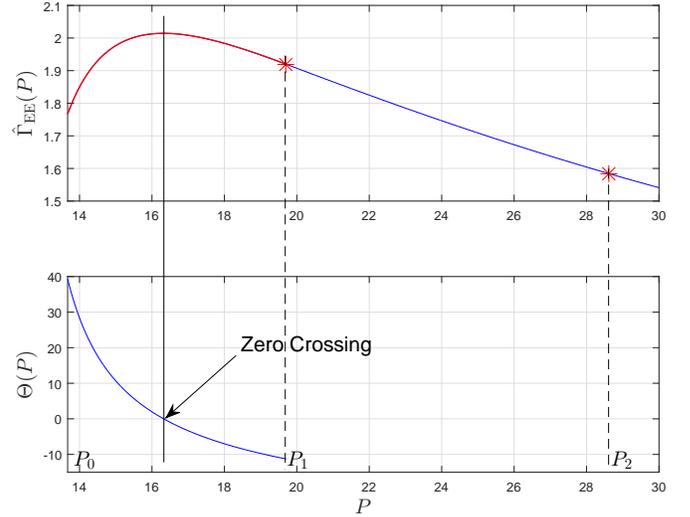}
  \caption{Search of $P^{opt}$ that maximizes the energy efficiency. Fixed bandwidth assignment.}
    \label{fig:EE}
\end{figure}

Fig.~\ref{fig:EE} reflects the accuracy of Algorithm \ref{alg:power} in Section~\ref{sec:EE_practical} for transmission at maximum energy efficiency.   In this example, the maximum energy efficiency is within interval $(P_{0}, P_{1}]$.  Therefore, only $\Theta(P)$ within this interval needs to be calculated and the zero-crossing point corresponds to $P^{opt}$ that maximizes $\hat{\Gamma}_{\mathrm{EE}}$.

\begin{figure}[t!]
  \centering
    \includegraphics[width=0.51\textwidth]{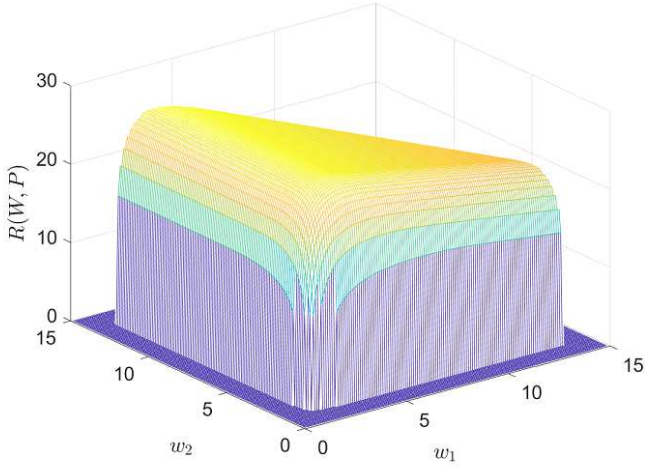}
  \caption{Sum information rate with arbitrary bandwidth assignment.  $W=10$, $P=20$.}
    \label{fig:R_w1w2P}
\end{figure}

\begin{figure}[t!]
  \centering
    \includegraphics[width=0.53\textwidth]{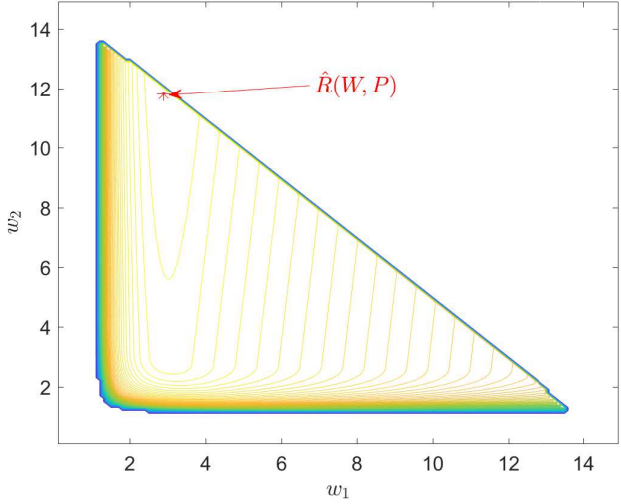}
  \caption{Contour plot of sum information rate.  Red star indicates $\hat{R}$ with joint optimal bandwidth assignment and transmit power allocation.}
    \label{fig:R_WP}
\end{figure}

Second, transmission efficiency is achieved with joint bandwidth assignment and transmit power allocation.  We randomly generate $g_{k}$ from $(0,10]$ and $\check{r}_{k}$ from $(0,10]$.  
Fig.~\ref{fig:R_w1w2P} shows the concave envelope of the sum information rate $R(W,P)$ with $W=10$, $P = 20$, and various bandwidth assignments.  Fig.~\ref{fig:R_WP} is a contour plot and the red star indicates the peak $\hat{R}(W,P)$ with joint optimal bandwidth assignment and transmit power allocation.  The calculation of $\hat{R}(W,P)$ follows \eqref{eq:max_rate_joint} in Section~\ref{sec:joint_optimization}.


\begin{figure}[t!]
  \centering
    \includegraphics[width=0.48\textwidth]{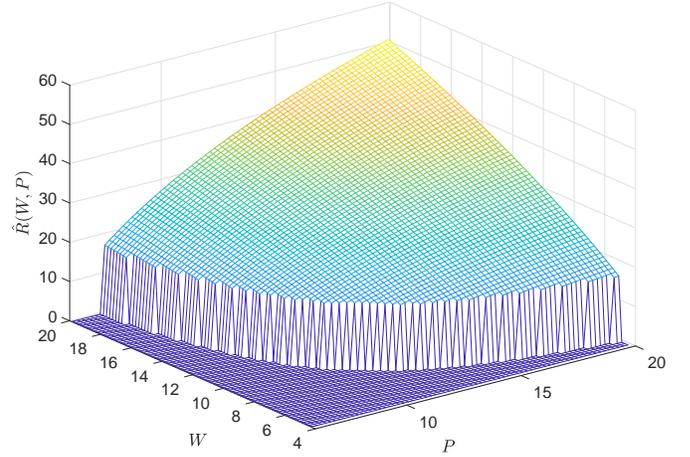}
  \caption{Sum information rate $\hat{R}(W,P)$ with joint optimal bandwidth assignment and transmit power allocation.}
    \label{fig:R_opt}
\end{figure}

\begin{figure}[t!]
  \centering
    \includegraphics[width=0.48\textwidth]{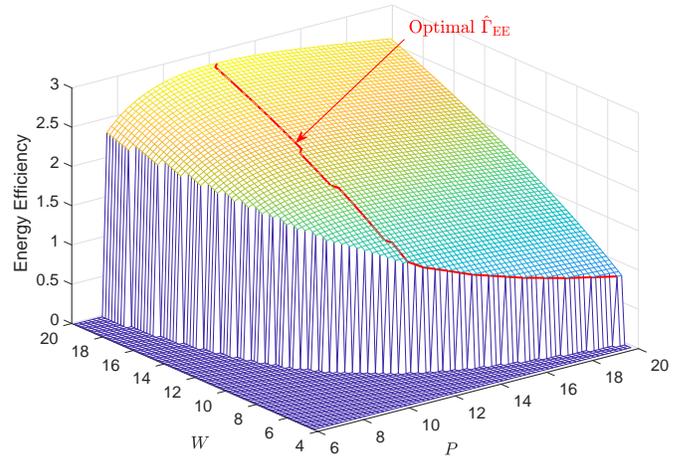}
  \caption{Energy efficiency of the multiuser transmission.  Solid red curve indicates $\hat{\Gamma}_{\mathrm{EE}}(P^{opt})$.}
    \label{fig:EE_opt}
\end{figure}

\begin{figure}[t!]
  \centering
    \includegraphics[width=0.48\textwidth]{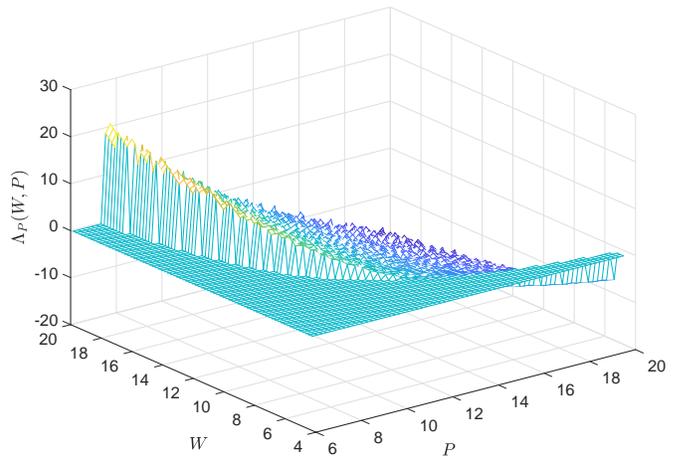}
  \caption{Sign indicator $\Lambda_{P}(W,P)$ of the efficiency derivative.  The zero-crossing curve corresponds to $\Lambda_{P}(W,P^{opt})$.}
    \label{fig:LamP_opt}
\end{figure}

As $W$ and $P$ change, Fig.~\ref{fig:R_opt} shows the sum rate $\hat{R}(W,P)$ with joint optimal bandwidth assignment and transmit power allocation.  $\hat{R}(W,P)$ increases as $W$ or $P$ increases.  Fig.~\ref{fig:EE_opt} shows the energy efficiency $\Gamma_{\mathrm{EE}}$.  For various $W$'s, the solid red curve indicates the peak $\hat{\Gamma}_{\mathrm{EE}}(P^{opt})$ that is found with the derivative-based method in Section~\ref{sec:combined_efficiency}.    $\hat{\Gamma}_{\mathrm{EE}}(P^{opt})$ increases as $W$ increases, and the maximum is at $W = W_{M}$.
Fig.~\ref{fig:LamP_opt} shows the sign indicator $\Lambda_{P}(W,P)$ of the efficiency derivative.  The zero-crossing curve corresponds to the total transmit power $P= P^{opt}$.  



\section{Conclusions}
\label{sec:conclusion}

Over parallel frequency channels, the problem of energy-efficient multiuser transmission is formulated with the bandwidth and transmit power constraints and the user-specific rate requirements.   The nominal carrier frequency of the channel for each receiver is pre-determined, and the channel frequency response is constant and known to the transmitter.  With fixed bandwidths of the parallel channels, a practical method is developed to find the total transmit power along with its optimal user allocation that maximizes the energy efficiency.   The optimal power allocation follows a multi-level water-filling solution.  With flexible channel bandwidths, joint bandwidth assignment and power allocation are developed. To maximize the energy efficiency, all the available bandwidth is used for multiuser transmission, and the optimal total transmit power $P^{opt}$ is found with an effective method.   The optimal solution of resource allocation is unique, and it guarantees the minimum rate requirement of each user while dedicating the excess spectrum and power resources to the one user with the best channel quality.

\appendices

\section{Proof of Proposition 1}
\label{ap:proposition1}

Problem $\mathcal{P}_{2}$ is a convex optimization problem.  Using the Karush-Kuhn-Tucker (KKT) conditions, we have ($\forall k \in \mathcal{K}$)
\begin{eqnarray*}
\check{p}_{k} \leq p_{k}, \,\,\,
\sum_{k \in \mathcal{K}} p_{k} &=& P \\
\lambda_{k} \geq 0, \,\,\, 
\lambda_{k} (\check{p}_{k} - p_{k}) &=& 0 \\
- \frac{w_{k}g_{k}}{w_{k}+ p_{k}g_{k}} - \lambda_{k}  + \mu  &=& 0.
\end{eqnarray*}
As $\lambda_{k}$ acts as a slack variable, it can be eliminated that yields
\begin{eqnarray}
\mu &>& \frac{w_{i}g_{i}}{w_{i} + p_{i}g_{i}}~~~\mathrm{and}~~~p_{i} = \check{p}_{i}, \,\,\, \forall i \in \mathcal{I} \subseteq \mathcal{K} \label{eq:mu_inequal} \\
\mu &=& \frac{w_{j}g_{j}}{w_{j} + p_{j}g_{j}}, \,\,\, \forall j \in \mathcal{K} \setminus \mathcal{I}. \label{eq:mu_equal}
\end{eqnarray}
Set $\mathcal{I}$ is the set of users that are transmitted to with their minimum required transmit power. Therefore, the optimal transmit power allocation is given by
\begin{eqnarray}
\hat{p}_{i} &=& \check{p}_{i}, \,\,\, \forall i \in \mathcal{I} \\
\hat{p}_{j} &=& w_{j} \left( \frac{1}{\mu} - \frac{1}{g_{j}} \right), \,\,\, \forall j \in \mathcal{K} \setminus \mathcal{I}.
\end{eqnarray}
With the total power constraint, we have
\begin{equation}
\frac{1}{\mu} = \frac{P - \sum_{i \in \mathcal{I}}\check{p}_{i} + \sum_{j \in \mathcal{K} \setminus \mathcal{I}}w_{j}/g_{j}}{\sum_{j \in \mathcal{K} \setminus \mathcal{I}}w_{j}}
\end{equation}
These equations can be combined as
\begin{equation} 
\hat{p}_{k} = \check{p}_{k}+ w_{k}\left( \frac{1}{\mu} - \alpha_{k} \right)^{+}, \,\,\, \forall k \in \mathcal{K} 
\end{equation}
where
\begin{equation}
\alpha_{k} = \frac{1}{g_{k}} + \frac{\check{p}_{k}}{w_{k}}.
\end{equation}

The maximum sum information rate with the optimal power allocation is
\begin{equation}
\hat{R} = \sum_{i \in \mathcal{I}}\check{r}_{i} + \sum_{j \in \mathcal{K} \setminus \mathcal{I}} w_{j} \log_{2} \left( \frac{g_{j}}{\mu} \right) .
\end{equation}

\section{Proof of Proposition 2}
\label{ap:proposition2}

Since $r_{k}$ is strictly increasing with $p_{k}$ and the optimal $\{\hat{p}_{k}\}$ is the water-filling solution, the maximum sum rate and the allocated transmit power for each user are nondecreasing in $P$. Consider the derivative of $\hat{R}(P)$ for $P \geq P_{0}$.  Increasing the total power by $\Delta P$, if Set $\mathcal{I}$ has no change, the limit is
\begin{eqnarray}
\lefteqn{\lim_{\Delta P \rightarrow 0^{+}}\frac{\hat{R}(P + \Delta P) - \hat{R}(P)}{\Delta P} } \nonumber \\ 
&=& \lim_{\Delta P \rightarrow 0^{+}} \frac{1}{\Delta P} \sum_{j \in \mathcal{K} \setminus \mathcal{I}} w_{j} \left( \log_{2}g_{j}\left( \frac{1}{\mu} \right)_{P+\Delta P} \right. \nonumber \\
&& \left. -  \log_{2}g_{j}\left( \frac{1}{\mu} \right)_{P} \right) \nonumber \\
&=& \lim_{\Delta P \rightarrow 0^{+}} \frac{1}{\Delta P} \sum_{j \in \mathcal{K} \setminus \mathcal{I}} w_{j} \log_{2}\left(1 \right. \nonumber \\
&& \left. + \frac{\Delta P}{P - \sum_{i \in \mathcal{I}}\check{p}_{i} + \sum_{j \in \mathcal{K} \setminus \mathcal{I}} w_{j}/g_{j}} \right) \nonumber \\
&=& \frac{1}{\ln 2} \cdot \frac{\sum_{j \in \mathcal{K} \setminus \mathcal{I}} w_{j}}{P - \sum_{i \in \mathcal{I}}\check{p}_{i} + \sum_{j \in \mathcal{K} \setminus \mathcal{I}} w_{j}/g_{j}}.
\end{eqnarray}
If Set $\mathcal{I}$ contracts, i.e., the $k$th user is excluded,\footnote{With multiple users simultaneously being excluded from Set $\mathcal{I}$ due to the increase in total transmit power by $\Delta P$, the limit can be calculated using a similar approach with $\Delta p_{k} = (w_{k}/\sum_{j \in \mathcal{J}'} w_{j}) \Delta P$.} 
the new sets are expressed as $\mathcal{I}' = \mathcal{I} \setminus \{k\}$ and $\mathcal{J}' = \mathcal{K} \setminus \mathcal{I}' = \mathcal{J} \cup \{k\}$.  With the optimal transmit power allocation, we have

\begin{eqnarray}
\left( \frac{1}{\mu}\right)_{P} &=&  \frac{P - P_{0} + \sum_{j \in \mathcal{J}}w_{j}\alpha_{j} }{\sum_{j \in \mathcal{J}}w_{j}} \nonumber \\
&=& \alpha_{k} \\
\left( \frac{1}{\mu}\right)_{P+\Delta P} &=& \frac{P + \Delta P - P_{0} + \sum_{j \in \mathcal{J}}w_{j}\alpha_{j} + w_{k}\alpha_{k}}{\sum_{j \in \mathcal{J}}w_{j}+w_{k}} \nonumber \\ 
&=& \alpha_{k} + \frac{\Delta P}{\sum_{j \in \mathcal{J}}w_{j} + w_{k}}.
\end{eqnarray}
Therefore,
\begin{eqnarray}
\lefteqn{\lim_{\Delta P \rightarrow 0^{+}}\frac{\hat{R}(P + \Delta P) - \hat{R}(P)}{\Delta P} } \nonumber \\ 
&=& \lim_{\Delta P \rightarrow 0^{+}} \frac{1}{\Delta P} \left[\sum_{i \in \mathcal{I}'} \check{r}_{i} + \sum_{j \in \mathcal{J}'} w_{j} \log_{2}g_{j}\left(\frac{1}{\mu} \right)_{P+\Delta P} \right. \nonumber \\
&& \left. - \sum_{i \in \mathcal{I}}\check{r}_{i} - \sum_{j \in \mathcal{J}} w_{j} \log_{2}g_{j}\left(\frac{1}{\mu} \right)_{P} \right] \nonumber \\
&=& \lim_{\Delta P \rightarrow 0^{+}} \frac{1}{\Delta P} \left[-\check{r}_{k} + w_{k} \log_{2} g_{k}\left(\frac{1}{\mu} \right)_{P+\Delta P} \right. \nonumber \\
&& \left. + \sum_{j \in \mathcal{J}} w_{j} \left( \log_{2} g_{j}\left(\frac{1}{\mu}\right)_{P+\Delta P} -  \log_{2} g_{j}\left(\frac{1}{\mu}\right)_{P}\right) \right] \nonumber \\
&=& \lim_{\Delta P \rightarrow 0^{+}} \frac{1}{\Delta P} \left[w_{k} \log_{2}\left(1 + \frac{\Delta P}{\alpha_{k}(\sum_{j \in \mathcal{J}} w_{j} + w_{k})} \right) \right. \nonumber \\
&& \left. + \sum_{j \in \mathcal{J}} w_{j} \log_{2} \left(1 + \frac{\Delta P }{\alpha_{k} (\sum_{j \in \mathcal{J}} w_{j} + w_{k})} \right) \right] \nonumber 
\end{eqnarray}
\begin{eqnarray}
&=& \frac{1}{\ln 2} \frac{w_{k}+ \sum_{j \in \mathcal{J}}w_{j}}{\alpha_{k} (\sum_{j \in \mathcal{J}} w_{j} + w_{k})} \nonumber \\
&=& \frac{1}{\ln 2} \frac{1}{\alpha_{k}} \nonumber \\
&=& \frac{1}{\ln 2} \cdot \frac{\sum_{j \in \mathcal{J}} w_{j}}{P - P_{0} + \sum_{j \in \mathcal{J}}w_{j}\alpha_{j}}. \label{eq:dR_dP}
\end{eqnarray}

The existence of the same limits indicates that $\hat{R}(P)$ is continuously differentiable in $P \geq P_{0}$.  Accordingly, the energy efficiency with optimal transmit power allocation $\hat{\Gamma}_{\mathrm{EE}}(P) = \hat{R}(P)/(P/\zeta + P_{C}) $ is continuously differentiable in $P$.   

As $P \geq P_{0}$, $d\hat{R}(P) / dP > 0$.  The second derivative is given by
\begin{equation}
\frac{d^{2}\hat{R}(P)}{dP^{2}} = -\frac{\sum_{j \in \mathcal{J}} w_{j}}{(P-P_{0}+\sum_{j \in \mathcal{J}} w_{j}\alpha_{j})^{2} \ln 2} < 0.
\end{equation}
Therefore, $\hat{R}(P)$ is strictly increasing and concave in $P \geq P_{0}$.

Denote the superlevel sets of $\hat{\Gamma}_{\mathrm{EE}}(P)$ as
\begin{equation}
S_{\beta} = \{P \geq \sum_{k \in \mathcal{K}} \check{p}_{k} \mid \hat{\Gamma}_{\mathrm{EE}}(P) \geq \beta \}.
\end{equation}
For $\beta \geq 0$, $S_{\beta} = \{P \geq \sum_{k \in \mathcal{K}} \check{p}_{k} \mid \beta P/\zeta + \beta P_{C} - \hat{R}(P) \leq 0 \}$.  $S_{\beta}$ is strictly convex in $P$ because $\hat{R}(P)$ is strictly concave in $P$.  Therefore, $\hat{\Gamma}_{\mathrm{EE}}(P)$ is strictly quasiconcave in $P$~\cite{ConvexOpt_Boyd}.  In addition, since $\hat{R}(P)$ is strictly concave in $P$, we have
\begin{eqnarray}
\lim_{P \rightarrow \infty} \hat{\Gamma}_{\mathrm{EE}}(P) = \lim_{P \rightarrow \infty} \frac{\hat{R}(P)}{P/\zeta + P_{C}} = 0.
\end{eqnarray}
It follows that $\hat{\Gamma}_{\mathrm{EE}}(P)$ is either strictly decreasing or strictly quasiconcave in $P \geq P_{0}$.

\section{Proof of Proposition 3}
\label{ap:proposition3}

With
$r = w \log_{2} \left(1 + q/w \right)$, 
where $q = pg > 0$,  the first and second derivatives of $r$ with respect to $w$ are
\begin{eqnarray*}
\frac{d r}{d w} &=& \log_{2} \left(1 + \frac{q}{w} \right) - \frac{q}{(q+w) \ln 2} \\
\frac{d^{2} r}{d w^{2}} &=& - \frac{q^{2}}{w(q+w)^{2} \ln 2}
\end{eqnarray*}
Let $\Phi(x) = \ln(1+1/x) - 1/(1+x), \forall x > 0$.  Because $\lim_{x \rightarrow \infty} \Phi(x) = 0$ and $d \Phi(x) /dx = -x/(1+x)^{2} < 0, \forall x>0$, we have  $\Phi(x) > 0, \forall x >0$.  The first derivative is positive $dr/dw > 0$ using $x = w/q$.  The second derivative is negative $d^{2}r/dw^{2} < 0$.  Therefore, $r$ is strictly increasing and concave in $w > 0$.  
It follows that
Problem $\mathcal{P}_{3}$ is a convex optimization problem with variables $\{w_{k}\}$ and $\{p_{k}\}$.  Using the KKT conditions, we have ($\forall k \in \mathcal{K}$)
\begin{eqnarray*}
\check{r}_{k} - w_{k} \log_{2}\left(1 + \frac{p_{k}g_{k}}{w_{k}} \right) &\leq& 0 \\
\sum_{k \in \mathcal{K}} w_{k} = W, \,\,\,
\sum_{k \in \mathcal{K}} p_{k} &=& P \\
\lambda_{k} \geq 0, \,\,\,
\lambda_{k} \left(\check{r}_{k} - w_{k} \log_{2}\left(1 + \frac{p_{k}g_{k}}{w_{k}} \right) \right) &=& 0 \\
(-1 -\lambda_{k}) \left( \ln \left(1 + \frac{p_{k}g_{k}}{w_{k}} \right) - \frac{1}{1 + \frac{w_{k}}{p_{k}g_{k}}}\right) + \nu_{1} &=& 0 \\
(-1 - \lambda_{k}) \frac{w_{k}g_{k}}{w_{k} + p_{k}g_{k}} + \nu_{2} &=& 0.
\end{eqnarray*}
Suppose that $\lambda_{i} > 0, \forall i \in \mathcal{I} \subseteq \mathcal{K}$ and $\lambda_{j} = 0, \forall j \in \mathcal{K} \setminus \mathcal{I}$.  Set $\mathcal{I}$ is the set of users that are transmitted to with their minimum data rates, i.e., $w_{i} \log_{2} (1 + p_{i}g_{i}/w_{i}) = R_{i}, \forall i \in \mathcal{I}$.  It follows that
\begin{eqnarray}
\nu_{1} > \Phi \left(\frac{w_{i}}{p_{i}g_{i}} \right)~~\mathrm{and}~~\nu_{2} > \frac{w_{i}g_{i}}{w_{i}+ p_{i}g_{i}}, && \forall i \in \mathcal{I} \label{eq:nu1nu2_ineq} \\
\nu_{1} = \Phi \left(\frac{w_{j}}{p_{j}g_{j}} \right)~~\mathrm{and}~~\nu_{2} = \frac{w_{j}g_{j}}{w_{j}+ p_{j}g_{j}}, && \forall j \in \mathcal{K} \setminus \mathcal{I} \nonumber \\ \label{eq:nu1nu2} 
\end{eqnarray}
where $\Phi(x) = \ln(1+1/x) - 1/(1+x), \forall x > 0$.  Because $d \Phi(x) /dx = -x/(1+x)^{2} < 0, \forall x>0$,
$\Phi(x)$ is a strictly decreasing function of $x > 0$.
If there are two distinct users $k, l \in \mathcal{K} \setminus \mathcal{I}$, \eqref{eq:nu1nu2} gives no positive solution for $w_{k}/p_{k}$ and $w_{l}/p_{l}$.  Therefore, Set $(\mathcal{K} \setminus \mathcal{I})$ contains at most one user to achieve the maximum sum rate $\hat{R}$.   The other $K-1$ users are in Set $\mathcal{I}$ which are transmitted to with their minimum data rates.  Suppose that user $k \in \mathcal{K} \setminus \mathcal{I}$.  From \eqref{eq:nu1nu2_ineq} and \eqref{eq:nu1nu2}, we have ($\forall i \in \mathcal{I}$)
\begin{equation}
\frac{p_{k}}{w_{k}} g_{k} > \frac{p_{i}}{w_{i}} g_{i} ~~~\mathrm{and}~~~
\frac{1}{g_{k}} + \frac{p_{k}}{w_{k}} < \frac{1}{g_{i}} + \frac{p_{i}}{w_{i}}
\end{equation}
If follows that $g_{k} > g_{i}, \forall i \in \mathcal{I}$.  The only user that is transmitted to with a data rate more than its required minimum data rate is the user with the best channel gain $g_{k} = \max\{g_{1}, g_{2}, \ldots, g_{K}\}$.  Suppose that the minimum rate of each user can be satisfied by the total bandwidth $W$ and the total used transmit power $P$. The bandwidth assigned and the transmit power allocated for the $k$th user are respectively $w_{k} = W - \sum_{i \in \mathcal{I}} w_{i} > 0$ and $p_{k} = P - \sum_{i \in \mathcal{I}} p_{i} >0$.  
The maximization problem of the sum information rate becomes the maximization of the achievable information rate of the $k$th user with the best $g_{k}$.  It is simplified as
\begin{eqnarray*}
\begin{array}{ll}
\begin{split}
\mathop{\mathrm{maximize}}_{\{w_{i}\}, \{p_{i}\} }
\end{split}  
& r_{k} = (W - \sum_{i \in \mathcal{I}} w_{i}) \\
& \,\,\, ~~~~~\cdot \log_{2} \left(1 + \frac{(P- \sum_{i \in \mathcal{I}} p_{i}) g_{k}}{W - \sum_{i \in \mathcal{I}} w_{i}} \right) \\
\mathrm{subject~to} & w_{i}\log_{2}\left(1 + \frac{p_{i}g_{i}}{w_{i}} \right) = \check{r}_{i}, \,\,\, \forall i \in \mathcal{I}.
\end{array}
\end{eqnarray*}

It can be shown that $r_{k}$ is a concave function of $\{w_{i}\}$ and $\{p_{i}\}$.  The maximum can be found by taking the gradient 
$\nabla r_{k} = 0$, which yields
\begin{equation}
\frac{1}{g_{i}} \left(2^{\frac{\check{r}_{i}}{w_{i}}} \left( \frac{\check{r}_{i}}{w_{i}} \ln 2 -1 \right) + 1 \right) = \psi, \,\,\, \forall i \in \mathcal{I}
\end{equation}
\begin{eqnarray}
\lefteqn{\frac{(P - \sum_{i}p_{i})g_{k}}{W - \sum_{i}w_{i}} + \psi g_{k} } \nonumber \\
&=& \left(1 + \frac{(P - \sum_{i}p_{i})g_{k}}{W - \sum_{i}w_{i}} \right) 
 \ln \left(1 + \frac{(P - \sum_{i}p_{i})g_{k}}{W - \sum_{i}w_{i}} \right)  \nonumber \\ \label{eq:calculate_phi}
\end{eqnarray}
where $\psi$ is the intermediate coefficient.  With $\psi$, the bandwidth and the transmit power of the $i$th user ($i \in \mathcal{I}$) can be solved as
\begin{eqnarray}
w_{i} &=& \frac{\check{r}_{i} \ln 2}{\mathcal{W}_{0}^{(i)} + 1} 
\\
p_{i} &=& \frac{(\exp(\mathcal{W}_{0}^{(i)} +1) -1)}{(\mathcal{W}_{0}^{(i)}+1)  }\cdot \frac{ \check{r}_{i} \ln 2}{g_{i}} \nonumber \\
&=& \frac{\psi g_{i}-1-\mathcal{W}_{0}^{(i)}}{\mathcal{W}_{0}^{(i)}(\mathcal{W}_{0}^{(i)} + 1)} \cdot \frac{\check{r}_{i} \ln 2}{g_{i}} 
\end{eqnarray}
where $\mathcal{W}_{0}^{(i)} = \mathcal{W}_{0}(\frac{\psi g_{i}-1}{e})$, and $\mathcal{W}_{0}(\cdot)$ is the principal branch of the Lambert W function.    From \eqref{eq:calculate_phi}, $\psi$ can be expressed as
\begin{equation}
\psi = \left(\frac{1}{g_{k}} + \frac{p_{k}}{w_{k}} \right) \ln\left(1 + \frac{p_{k}g_{k}}{w_{k}} \right) - \frac{p_{k}}{w_{k}}.  \label{eq:maximum_rk}
\end{equation}
Since the maximum achievable rate of the $k$th user is given by
\begin{equation}
\hat{r}_{k} = w_{k} \log_{2}\left(1 + \frac{p_{k}g_{k}}{w_{k}} \right)
\end{equation}
and $p_{k} = \frac{w_{k}}{g_{k}} (2 ^{\hat{r}_{k}/w_{k}} - 1)$, it follows that
\begin{equation}
\psi = \frac{1}{g_{k}} \left(2^{\frac{\hat{r}_{k}}{w_{k}}} \left(\frac{\hat{r}_{k}}{w_{k}} \ln 2 -1 \right) +1 \right).
\end{equation}
Solving for $w_{k}$, we have
\begin{equation}
w_{k} = \frac{\hat{r}_{k} \ln 2}{\mathcal{W}_{0}^{(k)}  +1}.
\end{equation}
Therefore, the maximum rate of the $k$th UE is given by
\begin{equation}
\hat{r}_{k} = \frac{w_{k} }{ \ln 2} \left(\mathcal{W}_{0}^{(k)}  + 1 \right)
\end{equation}
and the maximum sum information rate $\hat{R}$ is given by
\begin{eqnarray}
\hat{R} &=& \sum_{i \in \mathcal{I}} \check{r}_{i} + \hat{r}_{k} \nonumber \\
&=& \sum_{i \in \mathcal{I}} \check{r}_{i} + \frac{w_{k} }{ \ln 2} \left(\mathcal{W}_{0}^{(k)}  + 1 \right) .
\end{eqnarray}

\ifCLASSOPTIONcaptionsoff
  \newpage
\fi

\bibliographystyle{IEEEtran}
\bibliography{IEEEabrv,comp}








\end{document}